# Lost Data in Electron Microscopy

Nina M. Ivanova, Alexey S. Kashin, and Valentine P. Ananikov*

*N.D. Zelinsky Institute of Organic Chemistry, Russian Academy of Sciences, Leninsky Prospect 47, Moscow 119991, Russia;*
**e-mail: val@ioc.ac.ru, https://AnanikovLab.ru*

**ABSTRACT**

The goal of this study is to estimate the amount of lost data in electron microscopy and to analyze the extent to which experimentally acquired images are utilized in peer-reviewed scientific publications. Analysis of the number of images taken on electron microscopes at a core user facility and the number of images subsequently included in peer-reviewed scientific journals revealed low efficiency of data utilization. Up to around 90% of electron microscopy data generated during routine instrument operation remain unused. Of the more than 150 000 electron microscopy images evaluated in this study, only approximately 3 500 (just over 2%) were made available in publications. For the analyzed dataset, the amount of lost data in electron microscopy can be estimated as >90% (in terms of data being recorded but not being published in peer-reviewed literature). On the one hand, these results highlight a shortcoming in the optimal use of microscopy images; on the other hand, they indicate the existence of a large pool of electron microscopy data that can facilitate research in data science and the development of AI-based projects. The considerations important to unlock the potential of lost data are discussed in the present article.

## 1. INTRODUCTION

Modern progress in automated data processing, including the use of computer algorithms based on neural networks, greatly facilitated the solution of research tasks and fastened data analysis in chemistry, life science, nanotechnology and many other areas. Machine learning techniques are widely used to solve problems in synthetic and computational chemistry,[1–7] materials science[8–12] and catalysis.[13–18] The increasing use of machine learning approaches has made it possible to rapidly analyze large amounts of experimental data in different scientific fields. However, the issues of appropriate sharing[19–22] and storage[23–26] of scientific data, as well as realizing the potential for data reuse and rethinking,[27–30] become rather challenging. This is closely related to the questions of statistical significance and reproducibility of results obtained, the efficiency of using expensive and busy equipment, and the lack of comprehensive information in the scientific literature on negative results.

A vivid example of this type of scientific information is the data obtained from electron microscopy, a direct observation method now used to study the microstructure and nanostructure of materials.[31–33] Complex morphology, possible dynamic behavior and variations in micro- and nanostructures result in obtaining different images and a large overall number of microphotographs of a single sample. Each of the images taken may differ significantly from the others and is a separate source of scientifically valuable information. The use of computer-aided processing of electron microscopy data,[34–40] especially for dynamic systems,[41–46] in some cases allows comprehensive structural information to be obtained. However, despite the wealth of data available from electron microscopy experiment for a single sample, it seems that often only one or a few images confirming a particular hypothesis are used to illustrate a publication. There is a risk that the majority of the images taken in the experiment may remain unpublished. In view of this, the question of increasing the efficiency of the use of the results of microscopic studies and the problem of lost data in electron microscopy is of much importance.

In this article, we present an analysis of an array of electron microscopy data obtained during >10 years of operation at the center for structural analysis and the core user facility, and the fraction of electron microscopy images published in peer-reviewed journals to date. The consolidation and systematization of the scanning and transmission electron microscopy (SEM and TEM) data resulted in an array of approximately 152k initial images, 142k sets of parameters and 3577 images published in 292 articles. The results showed that more than 97% of the scientifically significant electron microscopy data were actually not published (lost for the development of science), demonstrating the critically low efficiency of data utilization and the need to revise and rationalize approaches to the use of microphotographs in scientific research. As a note, we have tried to focus on electron microscopy studies in chemical nanoscience, synthetic chemistry and catalysis, which are different from biological and environmental research. However, these areas have also been mentioned briefly to give a more comprehensive picture of the problem and possible solutions.

To the best of our knowledge, this is the first systematic study to quantitatively assess the extent of lost data in electron microscopy. By consolidating and analyzing over a decade's worth of real-world microscopy output, this work offers a data-driven perspective on how extensively experimental results are underutilized. This analysis not only reveals a substantial inefficiency in the dissemination of valuable scientific information but also opens up important discussions on research transparency, data reuse, and the cognitive biases that shape what gets published. The findings are broadly relevant across scientific domains where data-intensive methodologies are employed, and they underscore the need to rethink current practices in data management, publication standards, and the design of AI-driven discovery pipelines. Through this lens, the study contributes to a growing conversation on the cultural and infrastructural shifts required to enable more complete, equitable, and intelligent use of experimental data.

## 2. RESULTS AND DISCUSSION

### 2.1. General remarks and scope of the analysis

The first problem to be solved was to find a sufficiently comprehensive source of raw electron microscopy data for further analysis. Arrays collected directly on the microscope are a good source of data as they contain all the original micrographs taken during the measurements. Access to such third-party image repositories is usually limited, so we had to rely on the data we have access to, which was available in statistically significant quantities. This range of data included images obtained on scanning and transmission electron microscopes, while additional information such as X-ray microanalysis and diffraction data (selected area electron diffraction, SAED, or electron backscatter diffraction, EBSD) were excluded from consideration because of their relevance to only a limited range of samples and tasks. It should be noted that all the microphotographs included into analysis were free from significant artifacts, were not the result of imaging with incorrect or non-optimized parameters and were of good quality, allowing each of them to be used as a reliable source of structural information.

### 2.2. Data collection and preparation

In the primary step of the study, the initial data of the scanning and transmission electron microscopes installed in the center for structural analysis were collected and statistically analyzed. A total of 152097 images (403 GB of data) were prepared for analysis. The total number of SEM and TEM images was 119557 (143 GB of data) and 32540 (260 GB of data), respectively. Data processing scripts were written to allow analysis based on a number of different parameters (Figure 1).

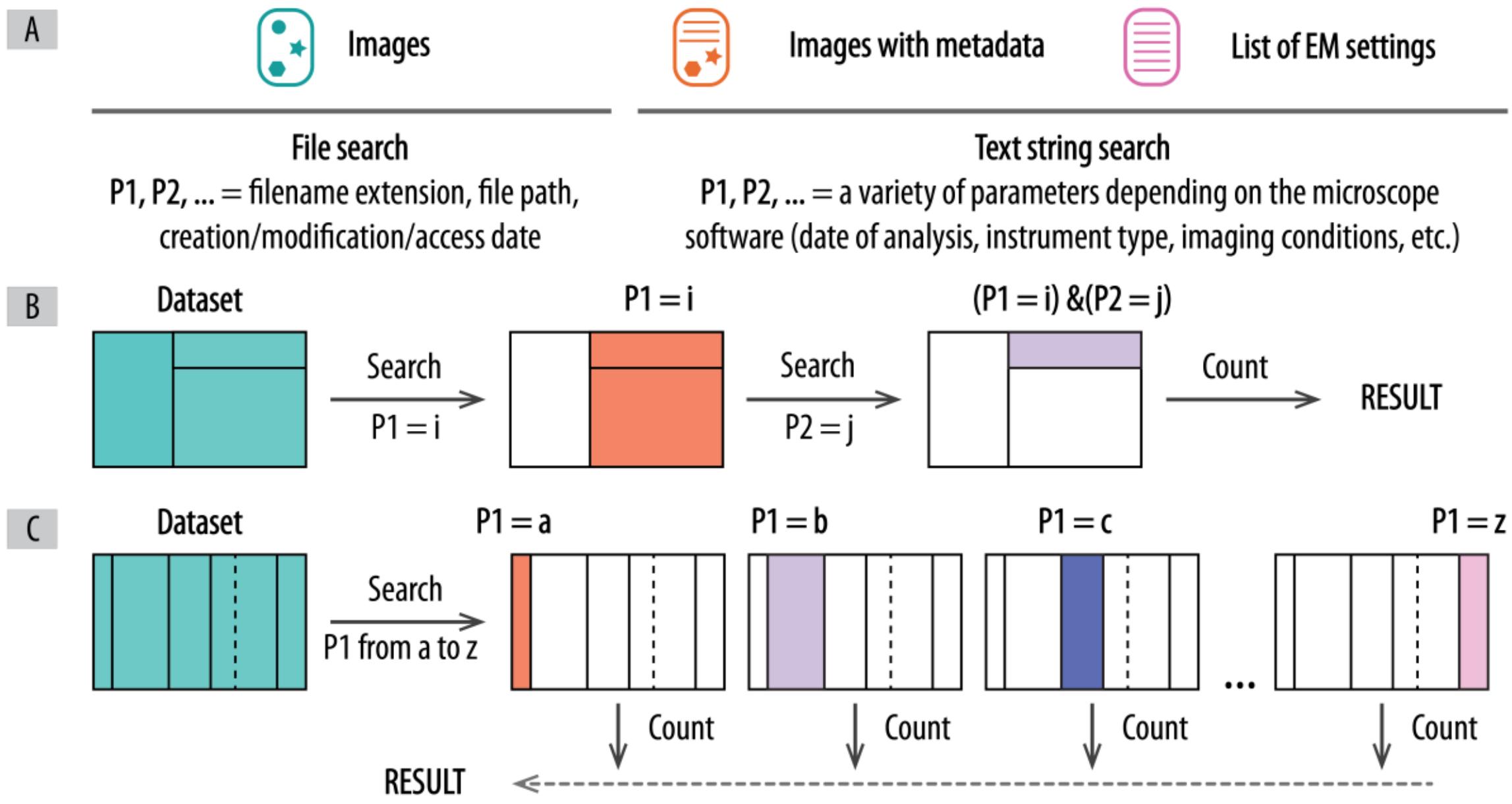

**Figure 1.** Different types of available electron microscopy data and corresponding strategies for their analysis (A). Two main algorithms used to analyze the array: sequential filtering of the array by several parameters (B) and sorting the whole array into different values of the same parameter (C).

The attributes of the files deposited in the data storage were chosen as the basic source of information for the analysis, and a file search was employed for array processing. The range of basic parameters relevant to the analysis included the file type and the date of its last modification, which made it possible to tentatively sort the electron microscopy images. A more complete analysis was possible using metadata with microscope characteristics and imaging parameters used or corresponding information stored in separate log files. It should be noted that the storage of the imaging parameters in text form actually proved to be less reliable and led to the loss of some data. A total of 141681 log files and image files with metadata were found in the archive. The range of stored imaging parameters available for analysis was greater than the number of significant file attributes and included, for example, microscope model, detector type, magnification and other characteristics (Figure 1A). In this case, the analysis was performed by a string search using regular expressions. The search was carried out using two algorithms. The first consisted of sequential filtering of the array by several parameters and then counting the number of remaining files (Figure 1B). This approach was used, for example, to determine the number of images taken in a given year on a microscope of a specific type using a particular detector (P1 = <year> & P2 = <microscope> &

P3 = <detector>). The second approach to analysis was to sort the whole array into different values of the same parameter and then count the number of files in each subarray (Figure 1C). This algorithm was used, for example, to sort microphotographs by the magnification value (P1 = <magnification-1, magnification-2, magnification-3,... >).

Time period of analysis was selected up to 2023, since publication time may sometimes be up to 1 – 2 years (during 1 year on average, but 2 years are not uncommon). For the images obtained in 2024 and 2025 currently it is not clearly possible to distinguish the published/unpublished status. Thus, the images obtained in 2024 and 2025 were excluded from the analysis.

### 2.3. Analysis of stored electron microscopy data

To obtain information about the quantitative composition of the entire archive of images, a string search was performed using three parameters: (1) year of image acquisition from 2011 to 2023, (2) microscope type (SEM or TEM), and (3) detector type (was used to separate the STEM images). Analysis of the data revealed a fairly steady increase in the number of images captured from year to year (Figure 2A). There is some variation in instrument use depending on the number of ongoing projects, the intensity of electron microscopy usage in each project, and hardware (re)configuration and repair, among other factors. Taking all these factors into account, the data represent a real load on a core user facility involved in a sufficiently large number of projects. On average, just over 10000 images are taken annually, of which 74% are SEM, 23% are TEM and 3% are STEM images. To some extent, this proportion of sample surface studies using SEM reflects the specificity of the tasks carried out at the core user facility, but it is interesting to obtain a more visual numerical expression of the nature of the objects studied. To solve this problem, we used a simple approach based on the analysis of image acquisition parameters, namely, on the analysis of the range of scale bar sizes used (Figure 2B).

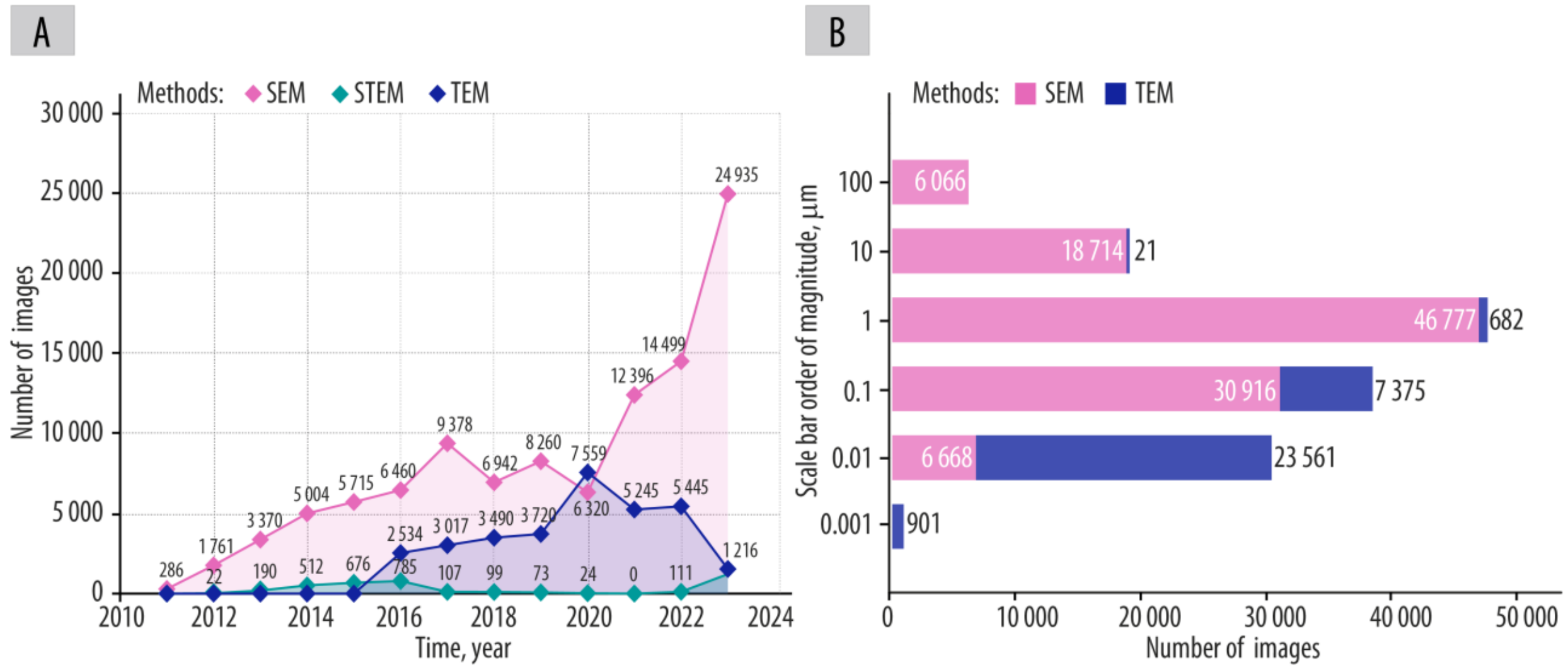

**Figure 2.** Summary of the composition of the electron microscopy image archive. Distribution of SEM and TEM images by year (A) and conditional scale bar size (B).

For this purpose, all images were grouped according to the magnification parameter. The magnification values were then sequentially converted to field of view (FOV) and scale bar length, which was conventionally assumed to be 10% of the FOV. As more than 80 different values of scale bar size were obtained, they were grouped according to the order of magnitude for clarity. The analysis showed that scanning electron microscopy, which operates in the micrometer range, and transmission electron microscopy, which focus mainly on nanoscale objects, together cover 6 orders of magnitude of characteristic sizes. It can therefore be concluded that the typical electron microscopy dataset contains a wealth of information on the morphology of a wide range of objects, from single nanoparticles to submillimeter assemblies and devices.

To check whether the content of the original electron microscopy dataset was fully reflected in the published articles, we analyzed the articles in peer-reviewed journals that included the data obtained at the core user facility.

### 2.4. Classification and analysis of published images

More than 1000 published articles mentioning electron microscopes installed in the user facility were analyzed in this study, with 292 publications containing electron microscopy images either in the main text or in the supplementary information. Expert analysis of the nature of the systems imaged allowed to tentatively group the articles into 5 broad categories: materials, catalysis, organic chemistry, ionic liquids and microscopic control (Figure 3).

The most voluminous category, “Materials”, is represented by 10 subcategories (Figure 3A). The most extensive subcategory includes membranes and porous materials: micro- and mesoporous carbon matrices of various origins, metal-organic frameworks, sorbents, molecular sieves, porous polymeric materials, etc. (references to specific articles are omitted here and below to avoid redundancy). In addition, separate subcategories have been identified that include articles devoted to the preparation and characterization of composite and hybrid materials, soft materials, biomaterials, polymer materials, minerals and ceramics. Two subcategories of carbon-based materials have also been identified. The first category includes pure carbon materials such as carbon quantum dots, carbon nanotubes, granular activated carbon, etc. The second subcategory includes metallic particles on carbon supports, as these materials are of particular importance as catalysts for many chemical transformations. The last subcategory is nano-scale particles, which are divided into metallic and nonmetallic subtypes. However, the borders between the selected subcategories may intercept because the same publication can cover both types of nanoparticles.

The “Catalysis” topic is subdivided into heterogeneous catalysis and homogeneous catalysis plus nanocatalysis. Heterogeneous catalysis includes systems in which a solid material plays the role of a support and the transfer of catalyst particles between different phases is insignificant or absent. All other systems were attributed to homogeneous catalysis. This category included both classical catalysts based on soluble metal complexes and catalysts based on metal nanoparticles in dynamic equilibrium with the molecular phase.

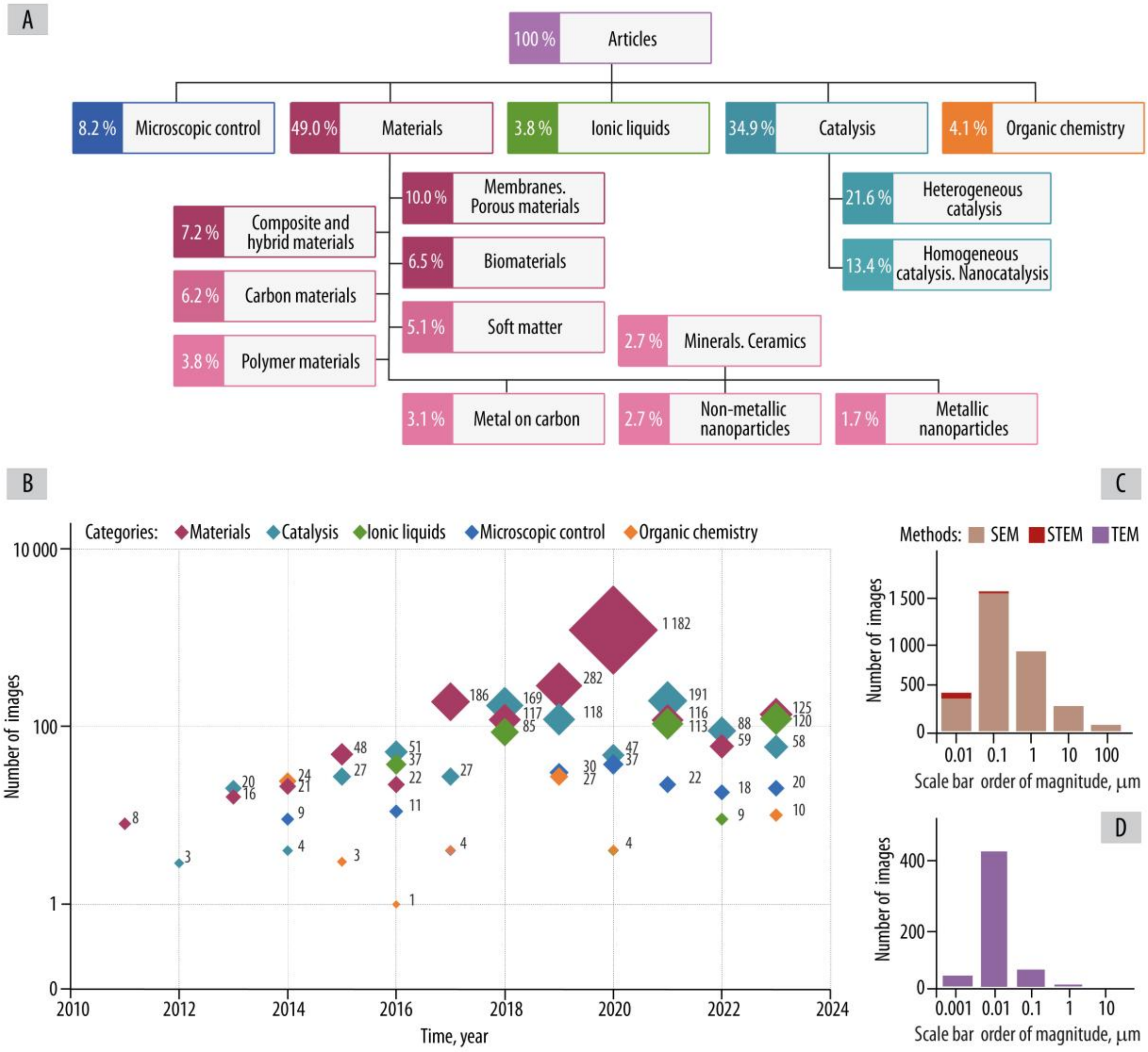

**Figure 3.** Summary of the published electron microscopy data. Percentage distribution of published articles by selected categories and subcategories (A) as well as distribution of the number of published images by article topic and year of publication (B). The characteristic sizes of the objects studied by S(T)EM (C) or TEM (D) in terms of the conditional scale bar size.

The analysis also identified a large group of articles that required the creation of specific subcategories: microscopic control, organic chemistry and ionic liquids. The first of these takes into account those cases where it is not a specific substance/material or its morphology that needs to be studied but rather a set of phenomena that occur under external stimuli. The increasing involvement of electron microscopy methods in new areas of research has led to a significant number of articles devoted to the study of organic chemical systems outside the classical fields of organic materials science, such as soft matter or polymer chemistry. In this regard, two distinct categories were identified: organic chemistry – electron microscopy observations related to solving problems in synthetic organic chemistry, and ionic liquids – examples of direct studies of liquid phase samples based on this class of liquid organic salts used as solvents.

During the whole period analyzed, 3577 microscopy images were published (Figure 3B), which averages over 12 images per article. It should be noted that this number is significantly overestimated due to the occurrence of occasional spikes in the number of images presented in papers or supplementary materials associated with the special cases of publication of large datasets. After only 8 articles containing more than 50 images each were removed, the total number of published microphotographs decreased to 1853 (for 284 articles), which corresponds to an average of 6.5 images per paper. As in the case of the array of experimental images, the published microphotographs were analyzed according to the order of magnitude of the scale bar. To construct the corresponding histograms, the field of view (FOV) values were extracted from each published image and processed in accordance with the abovementioned method. The results show that, in the case of SEM, the most valuable micrographs are those with a scale bar size corresponding to hundreds of nanometers (Figure 3C), which is an order of magnitude smaller than the typical value obtained by processing the entire dataset (Figure 2B). In the case of TEM, there is no difference between the acquired and published data (*cf.* Figure 2B and Figure 3D), and the most common scale bar size is on the order of 10 nm. Therefore, no valuable inconsistencies were found, making a significant portion of the acquired image array meaningful to researchers. Therefore, the small number of published images is not due to a lack of information provided by electron microscopy analysis but to the common practice of using microphotographs as illustrative material rather than as a self-sufficient result of structural research. It is worth noting that in many cases, this problem is solved by publishing the results of the statistical processing of a large number of images, but even this presentation of data results in the loss of significant morphological information compared to the original images.

### 2.5. Estimation of the data loss

The final step in the analysis was to compare the number of acquired and published electron microscopy images (Table 1) and to estimate the average number of electron microscopy images used in the publications as well as the amount of lost data. This part of the study is mainly based on data from the core user facility, whose collection and analysis methods are described in the previous sections of the paper. Also, as a reference, some publications containing electron microscopy images from third-party facilities were included in the analysis.

**Table 1.** Estimation of the amount of electron microscopy data used and lost.

| **Data source** | Core user facility | | | Third-party facilities | | | |
|---|---|---|---|---|---|---|---|
| **Data type** | SEM | TEM | TOTAL | SEM | STEM | TEM | TOTAL |
| **Number of acquired images** | 119557 | 32540 | 152097 | No data | No data | No data | No data |
| **Number of published articles** | 238 | 94 | 292 | 158 | 38 | 122 | 200 |
| **Number of published images** | 3054 | 523 | 3577 | 1624 | 274 | 818 | 2766 |
| **Average number of images per article** | 12.83 | 5.56 | 12.25 | 10.28 | 7.21 | 6.70 | 13.83 |
| **Published data, %** | 2.55 | 1.61 | 2.35 | – | – | – | – |
| **Lost data, %** | 97.45 | 98.39 | 97.65 | – | – | – | – |

The results in Table 1 show that SEM images (as well as STEM images, which fall entirely into this category, as they were recorded on a SEM instrument) taken at the core user facility appear in 238 publications, with the number of published microphotographs representing only 2.55% of the total SEM dataset. For TEM, this value is even lower and amounts to 1.61%. It should be noted that 40 articles included published images from both scanning and transmission electron microscopes, so the total number of articles was less than the simple sum of the number of articles for SEM and TEM separately. However, regardless of the type of instrument and the total number of images taken, the amount of data lost is on the order of 97-98%, and on average, only 2-3% of microphotographs are published.

Comparing the average number of images taken within the core user facility published in a single article with a similar parameter for third-party images derived from analysis of the content of 200 publications in peer-reviewed scientific publications on the topics of materials science, catalysis and nanotechnology showed that the value obtained for a particular source of microphotographs (12.25 based on both SEM and TEM images) correlated well with the general trends (13.83 based on SEM, STEM and TEM images). Therefore the results of statistical analysis described above can be considered reliable.

The results obtained demonstrate the extremely low efficiency of the use of electron microscopy data and, at the same time, indicate the great potential of using previously obtained experimental SEM and TEM results for further processing, for example, using modern AI technologies.

### 2.6. Solving the data loss problem

A detailed statistical analysis of the use of electron microscopy data described above revealed the existence of a data loss problem. At the same time, the identified features of the acquisition, storage and publication of electron microscopy data allowed to formulate a number of recommendations for the rational use of microphotographs in scientific research (Figure 4).

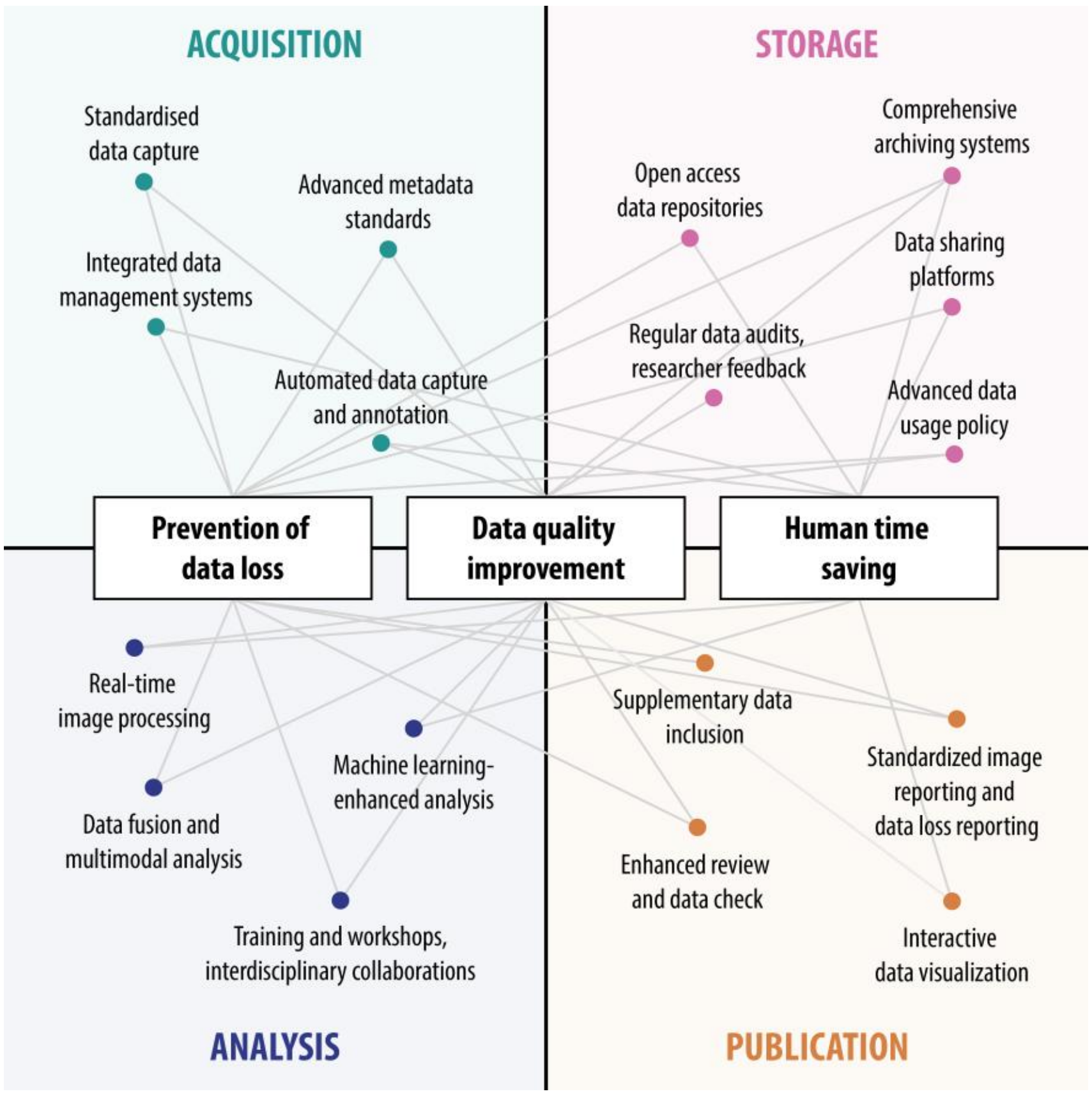

**Figure 4.** Schematic representation of summary of suggestions focused on improving electron microscopy data management, analysis and reporting to avoid the problem of data loss.

All the proposed solutions can be divided into several categories according to the stages of electron microscopy-aided research: image acquisition, storage, analysis and publication in scientific journals. In addition to direct approaches that make large amounts of data available, the ways to optimize the handling of raw data, to get more useful information from the same amount of electron microscopy images, and to optimize the time spent by researchers are also presented. It is worth mentioning that human time factor is of particular importance, as it is often a lack of time that leaves raw electron microscopy data unprocessed and therefore unsuitable for publication in peer-reviewed journals.

The standardization of image acquisition conditions, the most complete possible archiving of the conditions of electron microscopy experiments, and the careful cataloguing of images at the acquisition stage will lay the foundation for the creation of universal databases that can be used by researchers in different fields without the need to reproduce experiments on readily available or familiar equipment. The automation of these processes will significantly increase the efficiency of the researchers' work, which will contribute to a faster filling of the databases.

Electron microscopy data storage systems can be made publicly available. Capacity of modern web servers allows large amounts of data to be stored at minimal cost, making it possible to host micrographs on data sharing platforms. In addition, it is worth noting that open databases can also become discussion platforms to improve the quality of the data and respond to the current needs of researchers. Impressive efforts are currently being made to collect the thousands or even millions of microphotographs, including electron microscope images, and to share them between researchers in different scientific fields.[47] In particular, the introduction of a number of databases, e.g. the Image Data Resource (IDR, more than 14M images),[48] the Electron Microscopy Public Image Archive (EMPIAR, about 2k entries)[49] and the Australian Antarctic Data Centre electron microscopy database 1995-2007 (about 17k images)[50] should be mentioned. The agreement on the format and common standards for data storage[51,52] has facilitated the handling and reuse of microscopy data in the bioimaging community.

Undoubtedly, qualitative and complete processing of the acquired images will improve the efficiency of the data use. Modern image analysis systems, including those based on machine learning algorithms, will greatly simplify the analysis and can provide an opportunity for rapid data processing. Combining electron microscopy images with additional data for the same samples obtained by alternative methods, such as spectral techniques, can be a good way of extracting high quality scientific information from the original microphotographs. Collaboration between researchers from different disciplines in data analysis will greatly streamline the process and increase the proportion of scientific information generated that is suitable for widespread use.

Of course, there is also the need to improve the quality and amount of data published in traditional peer-reviewed scientific journals. The volume of data published can be significantly increased by making full use of the ability to attaching supplementary materials, and the way in which these materials are presented can also be improved in terms of faster access to the data and better visualization. The quality of the electron microscopy images presented will depend directly on the data publication policy, data presentation standards and the existence of the review and data check procedures.

We believe that considering these factors will improve the policy of electron microscopy data application in scientific research and allow more researchers to use this powerful and convenient technique in a rational way.

### 2.7. Importance of core user facility policy and impact on data loss

The core user facility analyzed in this study operates under a multiple-user access policy that provides researchers with direct and independent access to the electron microscopy equipment. In this model, researchers interested in using the microscopes undergo basic training, pass an equipment proficiency test, and are subsequently granted access to operate the equipment autonomously. Once certified, users are free to acquire microscopy data without limitations on the number of recorded images or experimental iterations. This flexible policy encourages exploration, trial-and-error optimization, and in-depth characterization of samples at the discretion of individual users and research groups.

The multiple-user access policy described here is quite common in research centers and corresponds to one of the typical modes of operation adopted by shared instrumentation facilities.

Such a policy is particularly advantageous for broadening access to advanced instrumentation and accelerating experimental throughput. However, it also results in the generation of large volumes of data with a substantial proportion remaining unpublished. In practice, many of the acquired micrographs are exploratory in nature, and although they may contain scientifically valuable information, they are often excluded from publications due to redundancy, selectivity, or time constraints during data processing and manuscript preparation.

It is important to note that the amount of lost data is likely influenced by the specific operational policies of user facilities. Facilities with different access models — such as centralized acquisition by trained staff, project-based scheduling, or pre-reviewed experiment planning — may produce significantly smaller or more curated datasets, potentially leading to a lower proportion of unused images. Therefore, the results and statistics reported in this work are closely tied to the policy of multiple-user access with unrestricted data generation, and should be interpreted in this context. Further comparative studies across various operational models may help to elucidate how user policies affect data retention and publication efficiency in scientific research environments.

In per-project analysis, for some unique samples with only a few areas of interest available, a much higher data usage rate may be observed. When only a few images are recorded in total (i.e., due to unique morphology of a certain single area, high cost of the sample with a small area, or other possible reasons) data usage/loss ratio may be ap-proximately 1:1. Thus, the amount of lost data may be reduced to around 50%. In overall, the number of such projects is very small, so it does not change the overall lost data value calculated for the entire dataset. As discussed above, the number of such projects may increase with different operational policies of user facilities.

## 3. CONCLUSIONS

This study presents the first systematic, data-driven quantification of lost data in electron microscopy, based on more than 150,000 micrographs acquired over a decade at a core user facility. A detailed comparison between the number of experimentally acquired images and those published in peer-reviewed scientific journals revealed that over 97% of recorded electron microscopy data remain unpublished and, therefore, largely unused in the scientific record. Specifically, only 2.55% of SEM and 1.61% of TEM images were found to be used in journal articles, indicating that the majority of image data generated by routine microscopy experiments are effectively lost.

One should not treat the specific value of 97% lost data as a universal metric, as the actual amount may vary depending on equipment usage policies, research practices, and institutional workflows; rather, the key conclusion is that the proportion of unused data is undeniably high and merits serious attention.

Given the variety of user policies and projects as mentioned above, it is reasonable to assume that the loss of 50–90% of data is not uncommon in electron microscopy. Even an estimate as low as 50% represents a significant amount of lost data, which has a strong impact on scientific research, equipment usage, and project economics.

It should be emphasized that the amount of lost data depends on the cost of the equipment and operation. For equipment/operations with a relatively small/regular cost, the amount of lost data may be larger since many images are taken and the time needed to record one image is rather small, as well as because scanning of large sample areas may take place. For high-cost equipment/operations, the amount of lost data may be significantly smaller due to limited availability, higher efficiency of use, or more time needed to record an image.

The analysis made in this study encompassed a wide range of research topics and imaging conditions and showed that the phenomenon of lost data is not due to poor data quality, but rather reflects a systemic inefficiency in data utilization. The findings provide the first quantitative evidence that vast amounts of high-quality scientific information are routinely discarded, highlighting a critical disconnect between data generation and its dissemination. The analysis of publications suggests that this pattern aligns with standard research practices and reflects a broader trend in scientific publishing and experimental workflows.

This work introduces a new perspective on the concept of "lost data" in experimental science and reveals an untapped resource with immense potential for data analysis, artificial intelligence training, and data-intensive research. It also emphasizes that the data loss rate may be strongly influenced by the operational model of user facilities, including policies on researcher access and data ownership.

The conclusions drawn here are not only related to microscopy but are broadly relevant to the design of institutional data policies, open science practices, and the strategic development of scientific infrastructure. Recognizing, quantifying, and addressing the problem of lost data is a necessary step toward improving research efficiency, enhancing reproducibility, and maximizing the return on investment in scientific instrumentation.

This study lays the groundwork for future efforts aimed at capturing, organizing, and repurposing unused microscopy data, and provides a model for similar assessments in other domains of experimental science.

To minimize the amount of data loss, the following key points should be considered:

(1) Employ reliable data acquisition protocols and use flexible, easily accessible storage solutions to facilitate data reporting, sharing, and reuse.

(2) Include all meaningful microscopy data in scientific publications, such as de-tailed imaging parameters and the results of seemingly "unsuccessful" experiments.

(3) Use automated data analysis to extract hidden structural information that could be useful for future research.

(4) Treat all high-quality images as sources of information for the future development of science.

(5) Use AI/ML tools to analyze all microscopy images obtained in the project and include the results in the published data domain.

Despite the large dataset analyzed and the comprehensive statistical approach, several limitations of this study should be acknowledged as well as directions for future re-search. The analysis was based on a single core user facility and primarily focused on chemical research applications of electron microscopy; therefore, extrapolation to other disciplines should be made with caution. The study also did not assess the scientific value of unpublished images directly, and for possible reasons for their exclusion on a per-project basis. Future work should explore qualitative aspects of unused data, user behavior in data selection, and facility-specific publication practices. Expanding this approach across multiple institutions and disciplines would help validate the generality of the findings and support the development of unified strategies for microscopy data retention, reuse, and sharing. Integration of AI tools for automated quality assessment and metadata enrichment may further transform how unused microscopy images are evaluated and incorporated into new research workflows. Analyzing the financial aspects of the data loss problem is also important. Clearly, the rational use of equipment can reduce the workload of microscopy devices, saving on maintenance and consumables purchases. Further research on this issue may focus on detailed economic studies and the sustainability of research practices.

## 4. METHODS AND EXPERIEMNTAL DATA PROCESSING

***Experimental details typical for the shared facilities operation.*** Electron microscopy imaging was carried out using a set of high-performance microscopes: scanning electron microscopes (Hitachi SU8000 and Hitachi SU8230 / Regulus8230) and a transmission electron microscope (Hitachi HT7700). These instruments provided capabilities for imaging in SEM (Scanning Electron Microscopy), TEM (Transmission Electron Microscopy), and STEM (Scanning Transmission Electron Microscopy) modes, depending on the nature and resolution requirements of the samples.

***Scanning electron microscopy (SEM).*** SEM imaging was performed primarily using the Hitachi SU8000 and SU8230 field-emission microscopes. Both instruments support high-resolution surface imaging and operate with accelerating voltages ranging typically from 0.5 to 30 kV. The working distance was adjusted between 3–15 mm depending on the desired depth of field and signal strength. Images were acquired using multiple detectors, including secondary electron (SE) detectors for topographic contrast and backscattered electron (BSE) detector for compositional imaging. Low-kV imaging (1–5 kV accelerating voltage) and beam deceleration technique (0.01–2 kV landing voltage) were frequently used for surface-sensitive analysis of non-conductive and beam-sensitive samples without the need for conductive coating. The resolution under high-voltage conditions reached sub-nanometer scales (<1.0 nm at 15 kV).

***Transmission electron microscopy (TEM).*** TEM imaging was performed on the Hitachi HT7700, an electron microscope with thermionic electron source (configuration with $LaB_6$ cathode was used, tungsten cathode can be installed as an option) optimized for materials characterization at relatively low accelerating voltages up to 120 kV. Samples were prepared as fine powders, thin films or ultramicrotomed sections with thicknesses generally below 100 nm to ensure sufficient electron transmission. TEM micrographs were usually recorded using bright-field (BF-TEM) mode. Selected Area Electron Diffraction (SAED) was used in specific cases to assess crystallinity and lattice spacing. The instrument allowed magnification from 1,000× to 800,000×, with typical imaging resolutions down to ~0.2 nm.

***Scanning transmission electron microscopy (STEM).*** STEM imaging was conducted using the SU8000 and SU8230 scanning electron microscopes equipped with the STEM detection systems, allowing high-resolution analysis in transmitted electron mode with nanometer and sub-nanometer resolution. The STEM mode was employed particularly for high-magnification imaging of nanoparticles, interfaces, and fine structural details. Bright-field (BF-STEM) and dark-field (DF-STEM) signals were collected depending on the contrast requirements. The pixel size, dwell time, and scan rate were optimized to balance resolution with beam damage, particularly for sensitive organic and hybrid materials.

***Imaging conditions and calibration.*** Across all instruments, imaging conditions such as accelerating voltage, beam current, aperture size, spot (pixel) size, magnification, and detector mode were optimized for each specific sample type. The microscopes were routinely calibrated using certified standard specimens to ensure dimensional accuracy and desired spatial resolution limit. Digital images were recorded in high-resolution formats (typically TIFF or JPEG) and were saved with corresponding metadata when supported by the software.

***Collection of the array.*** Electron microscopy images were obtained using equipment installed at the center for structural analysis and the core user facility for both scanning electron microscopy and transmission electron microscopy. The total amount of data collected was 403 GB or 152097 images and 109141 text files with parameters.

***Images filtering***. No manual selection or curation of image quality or project relevance was applied before statistical analysis. All micrographs meeting basic technical integrity (free of critical imaging artifacts) were included, thereby preserving the unbiased nature of the experimental dataset.

***Image acquisition parameters.*** Metadata files include detailed acquisition parameters such as acceleration voltage, working distance, detector mode, and magnification, which were used to reconstruct the imaging context and field of view for each micrograph.

***Analysis of the array.*** The data were analyzed using file search and string search tools (a string search was employed in the case of text files with parameters or images with metadata). The analysis was automated by using batch files. A summary of the search parameters and tasks is given in Table 2.

***Analysis of the published images.*** Peer-reviewed publications containing images from the abovementioned dataset were found in the Scopus, Web of Science and Google Scholar databases for the period from 2011 to 2023 inclusive. Full-text searches were performed using the microscope model, author affiliation and core user facility name. In addition, articles citing methodological and review papers written by core user facility staff were searched. Review articles and other types of publications with reused images (i.e., article translations) were excluded from the final set. In total, 292 publications were selected for further analysis. The published electron microscopy images were processed using ImageJ software.

The analysis of published data from third-party sources was based up to 2023 issues (the most recent issues for the analysis period selected in the article) of several journals in the fields of catalysis (ACS Catalysis, ACS), materials science (Advanced Materials Interfaces, Wiley), nanotechnology (ACS Nano, ACS) and microsciences (Small, Wiley). A total of 16 issues containing 474 articles were analyzed. Of these, 200 were selected that contained at least one electron microscopy image in the main text of the article or in supplementary materials.

**Table 2.** The list of parameters used to analyze the electron microscopy image dataset.

| Parameter | Search type | Formulated task/request |
|---|---|---|
| Type of the instrument | File search | Search for the image files (*.png, *.jpg, *.tif) in the tree folder transferred from a specific instrument's workstation |
| Year of image acquisition | File search, string search | File attribute search,<br>date search in DD.MM.YYYY format:<br><br>**dir** /t:w /s *.%Filetype% > <List><br>*Filetype = png, jpg, tif*<br><br>**find** /c /i ".%Year%" <List><br>*Year = 2011, 2012,…, 2023*<br><br>Search in parameter files (in metadata),<br>date search in MM/DD/YYYY format:<br><br>**findstr** /s /m "./%Year%" <Log file> > <List><br>*Year = 2011, 2012,…, 2023* |
| Type of the detector | String search | Search in parameter files (in metadata),<br>search for a specific value:<br><br>**findstr** /s /m " SignalName=%Detector%" <Log file> > <List><br>*Detector = SE, LA-BSE, HA-BSE, PDBSE, TE, BFSTEM, DFSTEM* |
| Magnification value | String search | Search in parameter files (in metadata),<br>search for a specific value:<br><br>**findstr** /s /m " Magnification=%Mag%" <Log file> > <List><br>*Mag = 20, 25, 30,…, 800000*<br><br>Search in parameter files (in metadata),<br>extraction of all values:<br><br>**findstr** /s /n " Magnification=" <Log file> > <List> |

***Experimental Workflow Overview***. A flowchart or schematic (optional) could illustrate the steps: data collection → storage audit → metadata parsing → statistical analysis → cross-referencing with publications. Details on each workflow stage is descried above in the text.

***Automated Processing.*** All batch processing and string searches were performed using automated scripts for minimizing human intervention and facilitating accurate processing.